\documentclass[12pt]{iopart}

\usepackage{amssymb}
\usepackage{graphicx}
\usepackage{dcolumn}
\usepackage{bm}
\usepackage{epsfig}
\newcommand{\beq}{\begin{equation}}
\newcommand{\eeq}{\end{equation}}
\newcommand{\beqa}{\begin{eqnarray}}
\newcommand{\eeqa}{\end{eqnarray}}
\newcommand{\ba}{\begin{array}}
\newcommand{\ea}{\end{array}}
\usepackage{color}
\begin{document}



\title[]{Long lived matter waves Bloch oscillations and dynamical localization by
time dependent nonlinearity management}
\author{Yuliy V. Bludov$^{1}$, Vladimir V. Konotop$^{1,2}$
and Mario Salerno$^{3}$}
\address{$^1$Centro de F\'{\i}sica Te\'orica e Computacional,
Universidade de Lisboa, Complexo Interdisciplinar,  Avenida
Professor Gama Pinto 2, Lisboa 1649-003, Portugal}
\address{
$^2$Departamento de F\'{\i}sica, Faculdade de Ci\^encias,
Universidade de Lisboa, Campo Grande, Ed. C8, Piso 6, Lisboa
1749-016, Portugal}
\address{$^3$Dipartimento di Fisica "E. R.
Caianiello" and Consorzio Nazionale Interuniversitario  per le
Scienze Fisiche  della Materia (CNISM),  Universit\'a di Salerno,
via S. Allende I-84081, Baronissi (SA), Italy}

\date{\today}

\begin{abstract}
We introduce a new method  to achieve long lived Bloch oscillations
and dynamical localization of matter wave gap solitons in optical
lattices. The method is based on time dependent modulations of the
nonlinearity which can be experimentally implemented by means of
the  Feshbach resonance technique. In particular, we show that the
width of the wave packet is preserved if time modulations of the nonlinearity
are taken proportional to the curvature
of the linear band spectrum which for most typical experimental
settings are well approximated by harmonic time modulations of
proper frequencies.
\end{abstract}
 \submitto{\JPB \rapid{}}
 \maketitle


Existence of non spreading localized waves (or solitons) is a
property of nonlinear systems which steams from the
balance of nonlinearity and dispersion.
In the case of Bose-Einstein condensates (BECs)
tightly bounded in two spatial dimensions, bright matter solitons
are known to exist when attractive two-body interactions
compensate the vacuum dispersion. These waves were experimentally
reported in Ref.~\cite{c:Kh2002}. An external periodic potential
generated by two counter propagating laser beams,
i.e. an optical lattice (OL), applied to a BEC gives rise to a
band-gap structure in the spectrum, introducing in this way
artificial dispersion.
In the effective mass approximation~\cite{c:KS}
one finds that signs of the atomic effective masses are different
at opposite edges of each finite gap. Under this condition the
properties of matter waves change significantly: a BEC with
repulsive inter-atomic interactions admits the existence of
stationary gap solitons  with chemical potentials inside
spectrum band gaps.
The experimental observation of these matter wave gap solitons was
reported  in ~\cite{c:Eiermann2004}. Due to broken translational
symmetry these solutions are intrinsically
localized, especially for high strength OLs, so that any attempt
to move them with the aid of an external force (such as, for
example, the acceleration of the OL), usually leads to their
destruction.

In order to control gap solitons and allow them to
move in real and in reciprocal spaces, one has to properly design
the nonlinearity in the system . Soliton existence
requires indeed that the effective nonlinearity and the effective
mass have opposite signs for all points in the Brillouin zone
\cite{c:KS}. In recent papers~\cite{c:BOnonl,c:DLnonl} it has been
shown that this condition can be satisfied  with the help of a
nonlinear optical lattice, i.e. with a spatial modulation of the
scattering length, and that it leads to the existence of Bloch
oscillations and dynamical localization of matter waves which are
long lived in presence of nonlinearity. A
nonlinear optical lattice imply a control of the scattering length
in space, a task which is more difficult to implement than the
modulation of the scattering length in time (this last can be
realized with external time dependent magnetic fields via the
usual  Feshbach resonance technique~\cite{Feshbach}). It is then
natural to ask if a time management of the nonlinearity can  lead
to the above long lived Bloch and dynamical oscillatory behaviors
in the nonlinear regime.

The aim of the present Letter is just devoted to
this. In particular we show for the first time, that long lived
Bloch oscillations and dynamical localization of a matter wave gap
soliton in an accelerated linear OL become possible in presence of
a properly designed time management of the nonlinearity. We remark
that simultaneous effects of linear OLs and nonlinear
time-dependent nonlinearities has been considered in literature in
various contexts. In the case of BECs this technique was used
mainly for the stabilization of higher dimensional solitons
~\cite{LattFeshbah-stab}, investigation of stable Feshbach
resonance-managed discrete matter-wave
solitons~\cite{LattFeshbah-reson} and generation of bright and
dark solitons~\cite{c:BraKon}. No applications of this technique
to nonlinear Bloch oscillations and dynamical localization
phenomena has been suggested so far. We remark that the
implementation of properly designed time dependent nonlinearities
for non destructive matter wave transport in OLs can be of
interest non only for BECs but also for nonlinear optics.

Let us start by considering a BEC described in the mean-field
approximation by the one-dimensional Gross-Pitaevski equation
(GPE) with a time-varying s-wave scattering length $a_s(T)$ and
cos-like OL accelerated in space according to the law $S(T)$:
\begin{eqnarray}
i\hbar \frac{\partial \Psi}{\partial T}=-\frac{\hbar^2}{2m}\frac
{\partial^2 \Psi}{\partial X^2}-V\cos\left[\frac{2\pi}{d}(X-S(T))\right]\Psi+
\frac{2\hbar^2a_s(T)}{a_\bot^2m}|\Psi|^2\Psi.
\label{eq:GPdim}
\end{eqnarray}
Here $m$ and $\Psi$ denote the boson mass and the macroscopic
wavefunction,
$V$ is the OL amplitude, $T$ and $X$
are the time and the longitudinal coordinate, respectively, $d$ is the
OL period, $a_\bot$ is the transverse trap size.
Eq.(\ref{eq:GPdim}) is obtained from the full three-dimensional
GPE in the case, when OL period $d$ is much bigger than
transversal trap size $a_\bot\ll d$ (for details of transition
see, e.g., \cite{c:BraKon}). Introducing the dimensionless moving
 spatial coordinate $x=\pi(X-S(T))/d$, time $t=TE_R/\hbar$
($E_R=\hbar^2\pi^2/(2md^2)$ is the recoil energy) and wavefunction
\begin{eqnarray}
\psi(x,t)=\frac{2d\sqrt{|a_s(0)|}}{a_\bot\pi}\Psi(X,T)
\exp{\left[i\frac{m}{\hbar}\frac{dS}{dT}(X-S(T))+ i\frac{m}{\hbar}
\int_0^T{\left(\frac{dS}{d\tau}\right)^2d\tau} \right]},\nonumber
\end{eqnarray}
we obtain the normalized GPE
\begin{eqnarray}
i \frac{\partial \psi}{\partial t}=-\frac{\partial^2 \psi}{\partial x^2}-v
\cos(2x)\psi+\gamma(t)x\psi+g(t)|\psi|^2\psi
\label{eq:GPdl}
\end{eqnarray}
with dimensionless OL amplitude $v=V/E_R$, time-dependent
nonlinearity $g(t)=a_s(T)/|a_s(0)|$ and external force
$\gamma(t)=(md/\pi E_R)d^2S/dT^2$, proportional to the
acceleration of the OL. In the absence of the nonlinearity
$g(t)\equiv 0$ and external force $\gamma(t)\equiv 0$, periodicity
of OL gives rise to band-gap structure of both spectrum $\varepsilon_n(q)$
and Bloch functions $\varphi_{nq}(x)$ of the linear problem
$\varepsilon_n(q)\varphi_{nq}=-d^2\varphi_{nq}/dx^2-v\cos(2x)\varphi_{nq}$,
where $n$ is the band number ( below we consider only the first band $n=1$ and
therefore the index $n$ is omitted) and $q$ is Bloch wavenumber. The
spectrum is periodic in the reciprocal space with the period $2$
(in the chosen units): $\varepsilon(q)=\varepsilon(q+2)$.

Under the influence of the
external force, soliton motion in the real and reciprocal spaces obeys
semiclassical laws for its center in real and reciprocal space $X$
and $Q$, correspondingly~\cite{c:BOnonl,c:DLnonl}:
\begin{eqnarray}
\dot{X}=\left.\frac{d\varepsilon}{dq}\right|_{q=Q}, \quad
\dot{Q}=-\gamma(t), \label{eq:semiclas}
\end{eqnarray}
where the overdot stands for the time derivative. When the
external force causes soliton motion, as it follows from
(\ref{eq:semiclas}), the position of soliton center $Q$ in
reciprocal space will be changed. At the same time the existence
of a gap  soliton solution of Eq.(\ref{eq:GPdl}) for a given
$Q(t)=Q_0-\int_0^t\gamma(\tau)d\tau$ ($Q_0$ is the initial
coordinate of the soliton in the reciprocal space) is determined
by the condition $M(Q(t))g(t)<0$ for all times $t$ (here
$M(q)=\left[d^2 \varepsilon(q)/dq^2\right]^{-1}$ is the soliton
effective mass). Thus, to prevent soliton destruction one should
vary the nonlinearity $g(t)$ in a manner to keep opposite sign
with respect the effective mass, (this assures the above existence
condition is satisfied for all $t$ and all $Q(t)$). Moreover, to
facilitate the soliton motion, one should minimize redistribution
of particles, keeping the soliton width to be constant (this can
be achieved by keeping the product of effective mass and effective
nonlinearity $g(t)\int_{-\pi}^{\pi}|\varphi_{Q(t)}(x)|^4dx$ to be
constant \cite{c:BOnonl}). For this it is sufficient to take
$g(t)$ of the form
\begin{equation}
g(t)= - \frac{|M(Q_0)|\int_{-\pi}^{\pi}|\varphi_{Q_0}(x)|^4dx}{M(Q(t))\int_{-\pi}^{\pi}|\varphi_{Q(t)}(x)|^4dx}.
\label{eq:cf-cond}
\end{equation}
In the following we apply the above nonlinear management scheme to
induce long lived nonlinear Bloch oscillations and dynamical
localization of a gap soliton in an accelerated OL.
\begin{figure}
  \begin{center}
   \begin{tabular}{c}
       \includegraphics{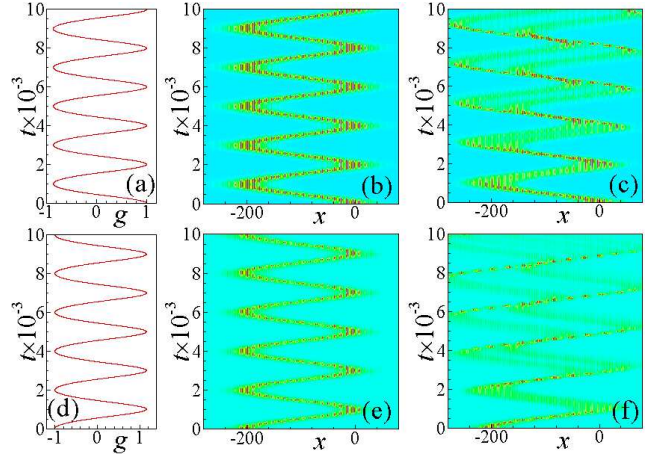}
   \end{tabular}
   \end{center}
\caption{(Color online) (a,d) Dependencies $g(t)$, providing
condition (\ref{eq:cf-cond}) for OL amplitude $v=3$, external
force  $\Gamma=-0.001$, and initial Bloch wavenumbers $Q_0=1$
(panel a), $Q_0=0$ (panel d). (b,e) Long-living Bloch
oscillations in BEC with time-varying nonlinearity, depicted in
panels (a) and (d), respectively. (c,f) Decaying Bloch
oscillations in BEC with the same parameters, as that of panels
b,e, respectively, except constant nonlinearity, which is $g(t)=1$
in (c) and $g(t)=-1$ in (f). The initial condition is the
stationary gap soliton with $E=-0.7324$, $Q_0=1$ (b,c) and
$E=-0.938$, $Q_0=0$ (e,f). All parameters are chosen the same as
in \cite{c:BOnonl}.} \label{fig:BO}
\end{figure}

Bloch oscillations of matter waves occur when BEC is subject to a
constant (dc) external force, i.e. for $\gamma(t)\equiv
\Gamma=$const. For the particular choice of the lattice amplitude
and the linear  force, the above  algorithm of designing the
temporal dependence (\ref{eq:cf-cond}) of the nonlinearity results
in curves depicted in Fig.~\ref{fig:BO}a,d for $Q_0=1$ and
$Q_0=0$, respectively. The gap-soliton dynamics under this $g(t)$
management is depicted in Fig.~\ref{fig:BO}b,e. Notice, that due
to proper choice of $g(t)$ Bloch oscillations become very regular
with the period and the spatial amplitude which coincide with the
values $T_s=2/|\Gamma|=2\cdot 10^3$ and
$X_s=[\varepsilon(1)-\varepsilon(0)]/(2|\Gamma|)\approx 32.5\pi$
predicted from Eqs.(\ref{eq:semiclas}). This behavior is in deep
contrast with the one depicted in Fig.~\ref{fig:BO}c,f for the
case of a constant nonlinearity, where the Bloch oscillations are
accompanied by the decay of the wave packet.

Similar results are obtained for the case of the dynamical
localization. In this case the motion of a soliton is turned on
under the action of a time-periodic external force
$\gamma(t)=\Gamma \cos\left(\Omega t\right)$. When the frequency
$\Omega$ is much smaller than the soliton energy $E$ (approximated by
the chemical potentials of the initial stationary state for the
initial $Q_0=0,1$) and at the same time fulfills the condition
$\Omega=\Gamma\pi/z_n$ with $z_n$ being the $n$-th zero of the
Bessel function $J_0(z_n)=0$~\cite{DK}, one expects dynamical
localization of the soliton, i.e. a regular oscillation in space
induced by the ac force, provided its existence is guaranteed for
all the energies belonging to the chose band. In~\cite{c:DLnonl}
this was achieved with help of spatially non uniform nonlinearity.
Now we design the time-dependent nonlinearity in order to obtain
the same effect.

\begin{figure}
  \begin{center}
   \begin{tabular}{c}
       \includegraphics{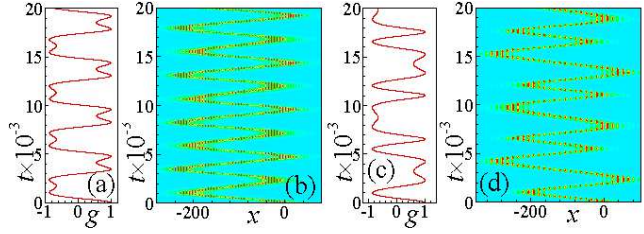}
   \end{tabular}
   \end{center}
\caption{(Color online) Dependencies $g(t)$, providing condition
(\ref{eq:cf-cond}) (panels a,c) and correspondent particle
densities $|\psi(x,t)|^2$ (panels b,d, respectively) for OL
amplitude $v=3$, initial Bloch wavenumber $Q_0=1$ and external
force (\ref{eq:tpef}) with parameters  $\Gamma=-0.001$, $t_0=500$
and frequencies $\Omega\approx 0.001306$ (panel a,b) and
$\Omega\approx 0.000569$ (panel c,d). All parameters are chosen in
accordance with \cite{c:DLnonl}. }
\label{fig:DL}
\end{figure}

Application of external force, which starts to oscillate at time $t=t_0$
\begin{equation}
\gamma(t)=\left\{\begin{array}{c}\Gamma,\quad t<t_0,
\\ \Gamma\cos\left[\Omega(t-t_0)\right], \quad t\ge
t_0,\end{array}\right.,
\label{eq:tpef}
\end{equation}
and time-dependent nonlinearity $g(t)$ (as depicted in
Fig.\ref{fig:DL}a,c) results in the dynamical localization of the
soliton, when its mean drift velocity becomes to be zero. This
results are obtained both for the oscillation frequency, which
corresponds to the first zero of Bessel function
(Fig.\ref{fig:DL}b) and for oscillation frequency, determined from
the second zero of Bessel function (Fig.\ref{fig:DL}d).

The numerical simulations, reported above, are performed for exact
numerical design of $g(t)$. However for typical values of $v\geq
1$ they are remarkably well reproduced using the dependence $g(t)$
which steams form the tight-binding approximation. Since in
typical experimental settings $v$ is of the order of several
recoil energies, expanding $\varepsilon(q)$ in the Fourier series,
we can consider only the two leading terms: $\varepsilon(q)=
\omega_0-\omega_1 \cos(\pi q)$. This approximation works well
already at $v\gtrsim 1$, as one can see form the Table I of
Ref.~\cite{AKKS}. For the particular case $v=3$ (for which our
above-mentioned numerical simulations are performed),
$\omega_0\approx -0.84$, $\omega_1\approx 0.102$. On the other
hand, in this approximation the integral in Eq. (\ref{eq:cf-cond})
becomes practically a constant, independent on $q$ and, hence, on
$t$~\footnote{\noindent The deviation from a
constant of the integral in (\ref{eq:cf-cond}), due to its
dpendence on $Q$, can be estimated in the Wannier basis as the
overlapping integral of Wannier functions of two adjacent sites,
this giving for $v=3$ an error of about 0.2 \%.} and the temporal
law (\ref{eq:cf-cond}) for nonlinearity $g(t)$ simplifies to
\begin{equation}
g(t)= -\cos\left(\pi Q(t)\right),
\label{eq:cf-cond2}
\end{equation}
i.e. it is uniquely fixed by the
dynamics of the wavepacket in the reciprocal space.

In this case of Bloch oscillations the semiclassical equations of
motion give $X(t)= X_0-\frac{\omega_1}{\Gamma}\cos(\pi
Q_0)\left[1-\cos(\pi\Gamma t)\right]$ and $Q(t)=Q_0- \Gamma t$
with $X_0$  the initial coordinate of the soliton. The temporal
dependence of the nonlinearity function then follows from
Eq.(\ref{eq:cf-cond2}) as $g(t)\approx -\cos(\pi
Q_0)\cos(\pi\Gamma t)$. At the same time for the dynamical
localization case time-dependent nonlinearity can be approximated
from Eq.(\ref{eq:cf-cond2}) as
\begin{equation}
g(t)\approx -\cos(\pi Q_0)\left\{\begin{array}{l} \cos (\pi\Gamma t),\quad
t<t_0,
\\ \cos\left(\pi\Gamma t_0+\frac{\Gamma}{\Omega} \sin\left[\Omega(t-t_0)
\right]\right), \quad t\ge
t_0,\end{array}\right.. \label{eq:g-local}
\end{equation}

To conclude, we have introduced a new method based on time
dependent modulations of the nonlinearity to achieve long lived
Bloch oscillation and dynamical localization of gap solitons in
OLs. The desired modulation is obtained directly
from the curvature of the underlying band structure and can be
experimentally implemented by means of time dependent external
magnetic fields via the Feshbach resonance technique. For
strengths of the OL greather than one recoil energy (e.g. for most
typical experimental settings), the shape of the time modulation
can be fixed in very simple manner using fields of proper
frequencies.

\vskip .5cm \noindent {\bf Acknowledgement} Y.V.B. acknowledges
support from FCT, Grant No. SFRH/PD/ 20292/2004. Cooperative work
is supported by the bilateral project within the framework of the
Portugal (FCT) - Italy (CNR) agreement.

\end{document}